\documentclass[5p, times]{elsarticle}
\usepackage{datetime}
\usepackage{amsmath}
\usepackage{amsfonts}
\usepackage{graphicx}
\usepackage{amssymb}
\usepackage{amssymb}
\usepackage{latexsym}
\usepackage{hyperref}

\def\abs#1{\left\lvert#1\right\rvert}

\begin{document}
\begin{frontmatter}

\title{Resonant kink-antikink scattering through quasinormal modes }
\author{Patrick Dorey}
\ead{p.e.dorey@durham.ac.uk}
\address{Department of Mathematical Sciences, Durham University, UK}
\author{Tomasz Roma\'nczukiewicz}
\ead{trom@th.if.uj.edu.pl}
\address{Institute of Physics, Jagiellonian University, Krak\'ow, Poland}

\begin{abstract}
We investigate the role that quasinormal modes can play in kink-antikink
collisions, via an example based on a deformation of the 
$\phi^4$ model.  We find that narrow quasinormal modes can store
energy during collision processes and later return it to the
translational degrees of freedom.  Quasinormal modes also decay, which
leads to energy leakage, causing a closing of resonance windows
and an increase of the critical velocity.  We observe similar
phenomena in an effective model, a small modification of the
collective-coordinate approach to the $\phi^4$ model.
\end{abstract}

%\maketitle
%\medskip

\end{frontmatter}

\section{Introduction}
It is well-known that kinks in 
nonintegrable models
such as the $\phi^4$ theory can
interact in a complicated way. One of the most
interesting features is the existence of a resonance structure in
kink -- antikink collisions
 \cite{Sugiyama:1979mi, Campbell:1983xu,Peyrard:1984qn}.  
During the initial impact (or `bounce'), 
oscillational modes can be excited, storing energy which on recollision can
be given back to the
translational modes of the kink and antikink. If the initial
velocities are right, a significant fraction of the energy is
returned, and kink-antikink pair can reseparate after one or more
further bounces, albeit with the loss of some energy to radiation
For other initial velocities less energy is returned, and the
kink-antikink pair annihilate, leading to a `fractal' structure of
nested escape windows \cite{Anninos:1991un, Goodman2005hbr}.

Such features were reported in many different models, including the double
sine Gordon model \cite{Peyrard:1984qn}, a coupled nonlinear Schr\"odinger
equation \cite{Yang}, and a two-component $\phi^4$ model
\cite{Halavanau:2012, Alonso-Izquierdo:2017zta,
Alonso-Izquierdo:2017gns}.
A collision of a kink with a suitable impurity \cite{Goodman2004impr,
Zhang1992, Goodman:2002dfct} or with a nontrivial boundary
\cite{Dorey:2017bdr, Dorey:2016bdr}
can also lead to resonant behaviour and a fractal structure.
Furthermore, a boundary collision can induce boundary decay
with the associated
creation of an extra kink or antikink,
resulting in a secondary resonant structure \cite{Dorey:2017bdr}.

For a long time it was thought that the existence of an
oscillational mode  of the kink was a necessary condition
for the formation of a resonant
structure.  More recently, it was shown that even in models such as the
$\phi^6$ theory, where kinks have no internal oscillational modes, a fractal
structure can still be observed, with modes trapped in the interval
between the
kink and antikink standing in for the localised modes \cite{Dorey:2011yw}. 
This new mechanism can be expected to lead to a fractal structure in
many cases of asymmetric kinks in models with different masses of
small perturbations around different vacua 
\cite{Gani2017, Demirkaya:2017euk, Simas:2017fjt, Menezes:2016mmh,
trom2017nrp1}.
Some efforts have also been made to reproduce the resonant structure of the $\phi^4$ theory in more realistic situations, such as graphene ribbons 
\cite{PershinGraphene}.

In this paper we exhibit yet another mechanism which
can lead to resonant scattering. Energy can
also be stored in narrow resonance modes, which in order to avoid 
confusion with the resonant structure will be called quasinormal modes
throughout this paper.
Quasinormal modes (QNM) are especially long-lived states which are
in some senses similar to oscillational modes, though
they satisfy purely outgoing boundary conditions and hence are not
normalisable.
They decay exponentially, losing energy due to their radiative tails. 

\subsection{Quasinormal modes}
Quasinormal modes play important roles both in quantum and classical
physics.  They satisfy purely outgoing wave boundary conditions,
breaking the hermicity of the Hamiltonian.  As a result, in quantum
physics,  they have complex energies $E=E_r+i\Gamma$.  The
imaginary part $\Gamma$ is responsible for exponential decay of the
state.  One of the earliest applications of this idea was in the
explanation of radioactivity:  a nucleus forms an effective potential
barrier which almost traps a particle, but which vanishes at
larger distances, allowing the particle to tunnel through it. QNMs can be
also seen as peaks of crosssections in scattering processes. 

Quasinormal modes are also often important in the classical evolution of
dynamical systems, and indeed that is the context where they were
first discussed \cite{Lamb}.
The long-time dynamics are governed by the poles of the Green's
function \cite{Kokkotas, bizon2, skyrmions}, with
the position of the pole determining the nature of the mode. 
Poles corresponding to real frequencies are normal oscillational modes
which in the linear approximation last infinitely long. Poles
corresponding to imaginary frequencies are unstable modes which grow
exponentially fast. 
The poles for complex frequencies describe QNM, which represent
decaying oscillations \cite{Kokkotas}. 
For massive fields so called threshold modes decaying according to
some power law can also dominate the long time dynamics
\cite{bizon2}.
One of the most surprising features of QNM is that their dynamics can
have nonlinear tails which start to dominate when the mode decays
below a certain amplitude \cite{bizon3}.

The most notable current
applications of QNMs are for signals of merging black holes. 
From gravitational wave measurements they can for example be used to
find the masses of the colliding black holes \cite{Anninos1998bh, Nakamura, Konoplya}.

\section{The model}
\subsection{Recalling the $\phi^4$ model}
In the following we limit our considerations to 1+1 dimensional
theories of a single scalar field:
\begin{equation}
 \mathcal{L}=\frac{1}{2}\dot\phi^2-\frac{1}{2}{\phi'}^2-U(\phi)
\end{equation}
The first example of the resonant scattering mechanism was
found for the $\phi^4$ theory, with the field theory (scalar) potential
\begin{equation}
U(\phi) = \frac{1}{2}(\phi^2-1)^2 \equiv W.
\end{equation}
The two vacuum configurations
$\phi(x)=\phi_{\pm}=\pm 1$
break the $\mathbb{Z}_2$ symmetry of the model; 
kinks and antikinks are stationary solutions which interpolate between
these vacua.
The static kink solution can be found from the BPS equation
\begin{equation}
 \phi_K'(x)=\sqrt{2U(x)}\,,
\end{equation} 
with a solution
\begin{equation}
 \phi_{K,\bar K}(x)=\pm\tanh(x)\,.
\end{equation} 
Small perturbations around the kink $\phi(x,t)=\phi_K(x)+e^{i \omega
t}\eta(x)$ satisfy the linearised equation
\begin{equation}
 -\eta''+V(x)\eta=\omega^2\eta,\qquad V(x)=
\biggl.U''(\phi(x))\biggr|_{\phi=\phi_K}
\end{equation} 
which has the form of a Schr\"odinger equation with a
`potential' for the linearised fluctuations given by
$V(x)$\footnote{to avoid confusion with the field theory potential
$U$, we will sometimes refer to $V$ as the linearised potential.}.
In this particular case this is the famous P\"oschl-Teller potential 
\begin{equation}\label{eq:potential1}
 V(x) = 4-\frac{6}{\cosh^2(x)}
\end{equation} 
and it supports two bound states, with frequencies
$\omega_0=0$ and $\omega_d=\sqrt{3}$. 
The first is the translational mode of the kink, while the second is
referred to as the oscillational mode.
The existence of this oscillational mode leads
to the resonance windows during kink-antikink collisions.
Some of the initial kinetic energy is stored in the oscillational mode,
which for appropriate (resonant) initial conditions can be given back to the
translational degrees 
of freedom in a subsequent recollision. The kink and antikink
can bounce multiple
times and either separate or end their existence as an oscillon. 
The resonant structure is very complicated, exhibiting fractal-like
 properties.  The model has been
studied extensively using both numerical and analytical methods. 
An effective model was introduced \cite{Sugiyama:1979mi,
Campbell:1983xu} which later was used in many variants and
approximations \cite{Anninos:1991un} and 
reproduced reasonably well both the fractal structure, and the critical
velocity above which no multibounce windows are observed.
However, it is important to note that the initial effective model
contained some errors, which were corrected in \cite{Weigel}.

\subsection{Designing the model}
Our aim is to study the influence of QNM on
collision scenarios similar to those known in the literature.  The
kink of the
standard $\phi^4$ model, defined above, does not have QNM in its
spectrum of small perturbations.  This is a rare feature, in this case
a consequence of the reflectionlessness nature of the linearised
potential for fluctuations about the $\phi^4$ kink.
Our strategy will be to modify the field theory potential $W$ so
as to turn the oscillational mode about the kink into a quasinormal
mode.  The linearised potential for the $\phi^4$ kink tends to
the asymptotic value $V(|x|\to\infty)=4$, meaning that waves with
frequencies below $2$ cannot propagate. However if at some distance
from the kink this
potential would decrease further, changing its asymptotic to
$V(|x|\to\infty)=m^2<4$,  waves with
frequencies below $2$ but above $m$ would become able to 
propagate.  In particular if $m<\sqrt{3}$
the oscillational mode could tunnel through the barrier and would
become a quasinormal mode.  We will use of this observation to design a
model for which the linearised potential for fluctuations about a
static kink is very similar to that of 
the $\phi^4$ model and yet its height
decreases as $\abs{x}\to\infty$. 

It is worth mentioning that having the linearised potential $V(x)$ one
can in principle
reconstruct the field theory potential $U(\phi)$ \cite{Bordag,skyrmions2}.
However, except some rare cases, the procedure gives a very
complicated potential which only can be found numerically, so we will
not adopt this approach. Instead, we look for
a field theory potential which is very similar to the
$\phi^4$ potential, $U\approx W$, when the field is far away from
either vacuum.  But when the field approaches one or other vacuum,
which will happen far from the kink, the behavior of the
potential should change. Recall that $V(x)=U''(\phi(x))$. For
$\phi=\pm 1$ the linearised potential is equal to $m^2$, which is the
squared mass of the scalar field. For the $\phi^4$ theory with our
normalisations, this mass is equal to $2$.  The second
feature which we want for our field theory potential is that its
second derivative around the vacuum $\phi=\pm 1$ would be $m^2 < 3$ to
allow the oscillational mode to tunnel through the barrier and to
become the QNM.
 
\begin{figure}
 \begin{center}
 \includegraphics[width=1\linewidth]{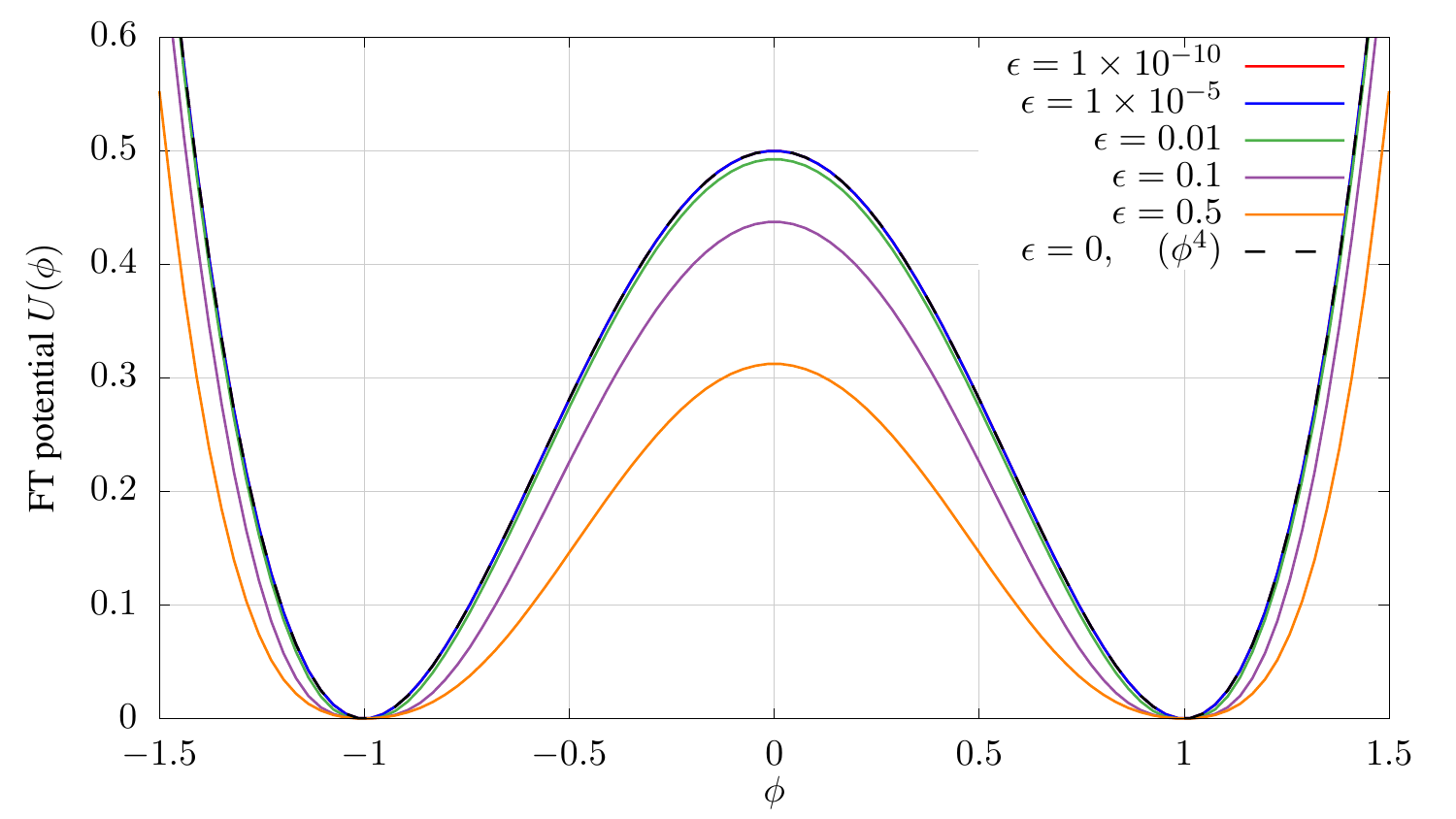}
 \caption{\small The field theory potentials $U(\phi,\epsilon)$
for various values of $\epsilon$.}
\label{fig:1:potentials}
\end{center}
\end{figure}

We have found one such family of field theory potentials to be
\begin{equation}
 U(\phi,\epsilon) = W+\frac{m^2-4}{4}\frac{\epsilon W}{W+\epsilon},
\end{equation} 
which for $\epsilon=0$ restores the standard $\phi^4$ potential. For $\epsilon>0$ the potential has a shape close to $\phi^4$, but near vacua (where 
$W\lesssim\epsilon$) it behaves as a field with mass $m$. Unless stated otherwise throughout the paper $m=1$.
Some examples of this potential for different values of $\epsilon$ are
shown in Fig.~\ref{fig:1:potentials}.

The $\phi^4$ kink approaches its vacuum as
\begin{equation}
 \phi_{\phi^4}(x)=\tanh x\approx 1-2e^{-2x}
\end{equation} 
For small values of $\epsilon$, 
the additional term in the potential becomes important
when $W\approx\epsilon$, which is for 
\begin{equation}
 x\approx-\frac{1}{4}\log(\epsilon/8).
\end{equation}
Beyond this point the approach to the vacuum changes.
The BPS equation (for $\phi=1-\xi$) takes the asymptotic form
\begin{equation}
 \xi'=-m\xi+\frac{m}{2}\xi^2-\frac{4-m^2}{m\epsilon} \xi^3+\mathcal{O}(\xi^4).
\end{equation} 
Note that the third term is singular in $\epsilon$. 
This means that in general the above expansion is not valid for $\epsilon=0$.
The singular term can be canceled only when $m=2$ which restores the
$\phi^4$ model.  For $\epsilon>0$ there is always such $\xi>0$ that
the third term can be neglected and the approach to the vacuum can be
found as:
\begin{equation}
 \xi(x)\approx \frac{2}{1+Ce^{mx}}\approx
2\left(\frac{e^{-mx}}{C}-\frac{e^{-2mx}}{C^2}+\frac{e^{-3mx}}{C^3}\right)+\cdots\,,
\end{equation} 
where $C$ is an integration constant depending on $\epsilon$.
When the expansion in $e^{-mx}$ is used the solution of the equation
including also the third term can be found. For 
$m=1$ it can be written as
\begin{equation}
 \xi(x)\approx 
a(\epsilon)e^{-x}-\frac{1}{2}a^2(\epsilon)e^{-2x}+
\frac{6+\epsilon}{4\epsilon}a^3(\epsilon)e^{-3x}+\cdots
\end{equation} 
where $a(\epsilon)>0$ is some constant, depending on the perturbation
parameter $\epsilon$,  
which can be estimated by matching the above solution with $\tanh(x)$
at the distance $x_0$.
The asymptotic of the linearised potential $V(x)$ changes from 
\begin{equation}
 V(x)\approx4-24e^{-2x}+\cdots
\end{equation} 
to 
\begin{equation}\label{eq:asspot}
 V(x)\approx 1-3ae^{-x}+\frac{3a^2(12-\epsilon)}{\epsilon}e^{-2x}+\cdots
\end{equation} 
Note that for small values of $\epsilon$ the coefficient standing
before $e^{-2x}$ is large and the third term can be larger than the
second until a certain distance is reached.

\begin{figure}
 \begin{center}
 \includegraphics[width=1\linewidth]{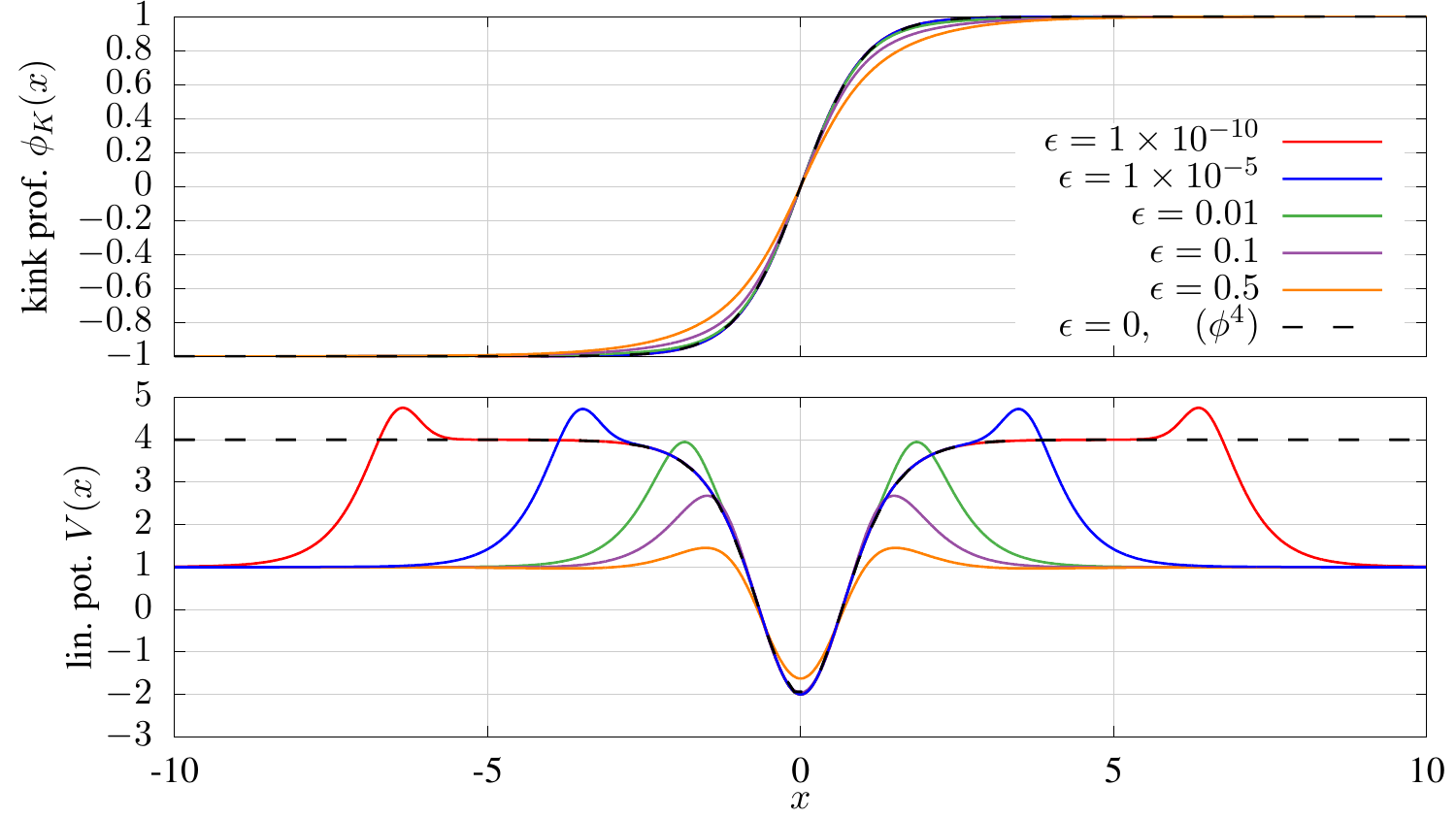}
 \caption{\small Profiles of the kinks and their linearised 
potentials.}\label{fig:2:kinks}
\end{center}
\end{figure}

Figure \ref{fig:2:kinks} shows the kink profiles and linearised
potentials $V(x)$ for various values of $\epsilon$. 
When $\epsilon$ is small, the form of $V(x)$ near to the centre of the
kink is close to that of the P\"oschl-Teller 
potential of the $\phi^4$ model. 
Far away from the kink, the linearised
potential, after a small bump, drops to 1.
Note that this behavior cannot be explained when only first two terms
of (\ref{eq:asspot}) are taken into account.
As mentioned earlier, for certain range of 
$x$ the third term with
$\epsilon$ in the denominator is larger than the second, matching the
fact that $V(x)$ appears to be larger than 1 in the asympototic region
in the plots on Figure \ref{fig:2:kinks}. Nevertheless,
we have checked that indeed $V(x)$ ultimately approaches 1 from below,
as would be expected from (\ref{eq:asspot}),
though this effect is small and only kicks in at larger values of $x$.

Hence the linearised potential has the property which we were looking
for:
for a certain distance it is almost equal to 4, and then it drops to 1.
Limited to a finite distance the potential has a bound state with a
frequency close to $\omega_d$. 
This mode can tunnel through the wall and leak to infinity. 
The parameter $m$ controls the asymptotic height of the potential and
$\epsilon$ controls the width of the barrier $x_0$. 
The larger the barrier, the more difficult the tunneling and the longer the
lifetime of the QNM. 
It is also worth noting that for $\epsilon>0$ the potentials do not
have any oscillational modes. 
Moreover the potentials are symmetric and no alignment of kinks and
antikinks can form a potential trap of the sort seen in the
$\phi^6$ model.
Therefore neither of the known mechanisms explaining resonant
structure is relevant for these models.

\begin{figure}
 \begin{center}
 \includegraphics[width=1\linewidth]{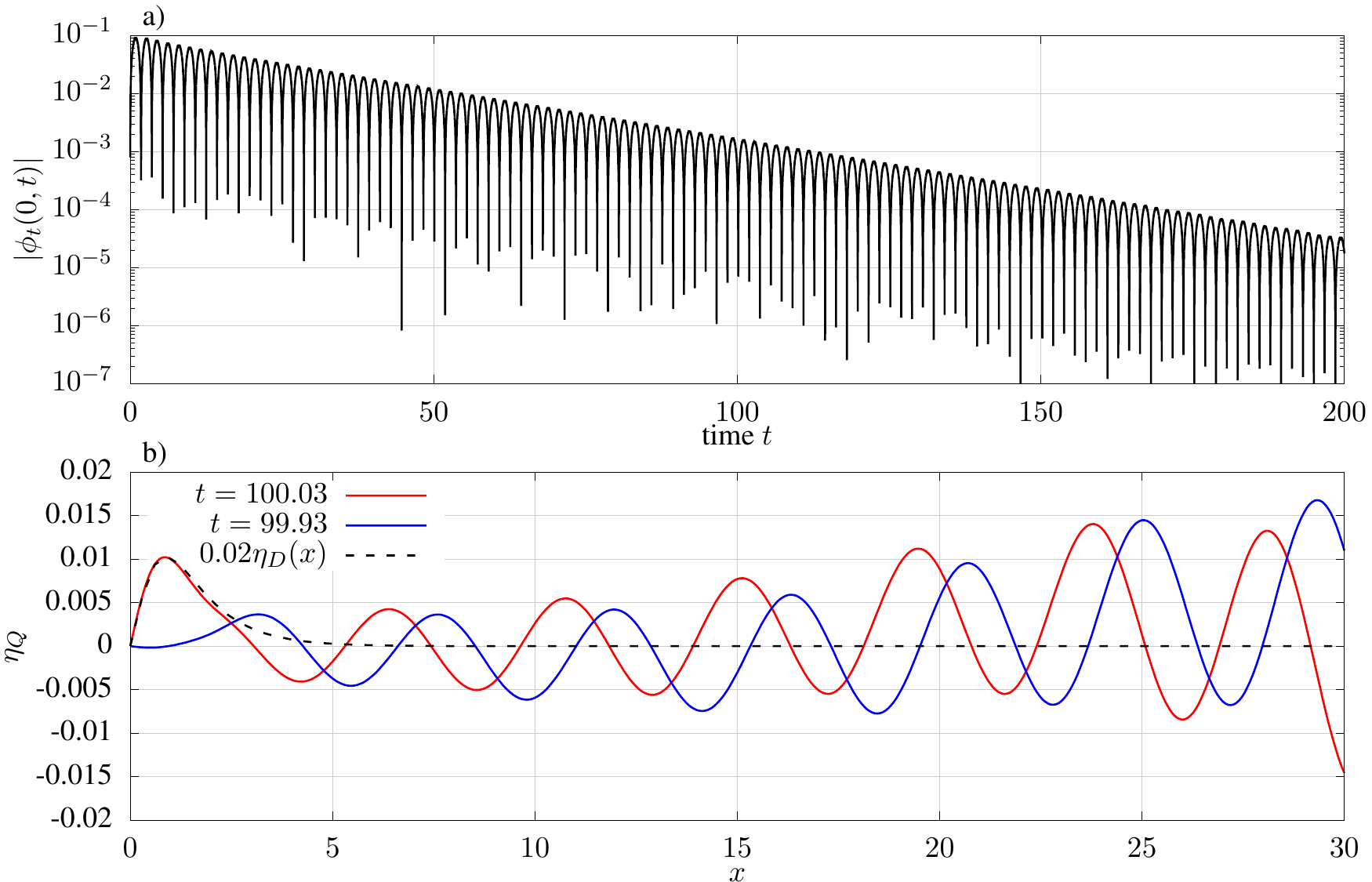}
 \caption{\small a) Exponential decay of the QNM for initial
conditions $\phi(x,0)=\phi_K(x)+0.1\eta_D(x)$, $\phi_t(x,0)=0$ in the
full nonlinear PDE. 
 b) Profile of the QNM for different times compared with the profile
of $\phi^4$ oscillational mode (dashed line) for linearised equation.
$\epsilon=0.01$. }\label{fig:QNMEvolution}
\end{center}
\end{figure}

Formally the QNMs satisfy the same linearised equation as 
oscillational modes.  
However the boundary conditions are different. 
For $\phi=\phi_K+e^{i\omega t}\eta(x)$ we require that far away from the kink 
\begin{equation}
\eta(x)\to e^{-ik|x|},\qquad k=\sqrt{\omega^2-m^2}.
\end{equation} 
This condition cannot be fulfilled for real frequencies, and complex
frequencies $\omega=\Omega+i\Gamma$ must be used. 
The imaginary part $\Gamma>0$, often referred to as the width, is
responsible for the exponential decay of the mode, as seen in Figure \ref{fig:QNMEvolution}. 

\begin{figure}
 \begin{center}
 \includegraphics[width=1\linewidth]{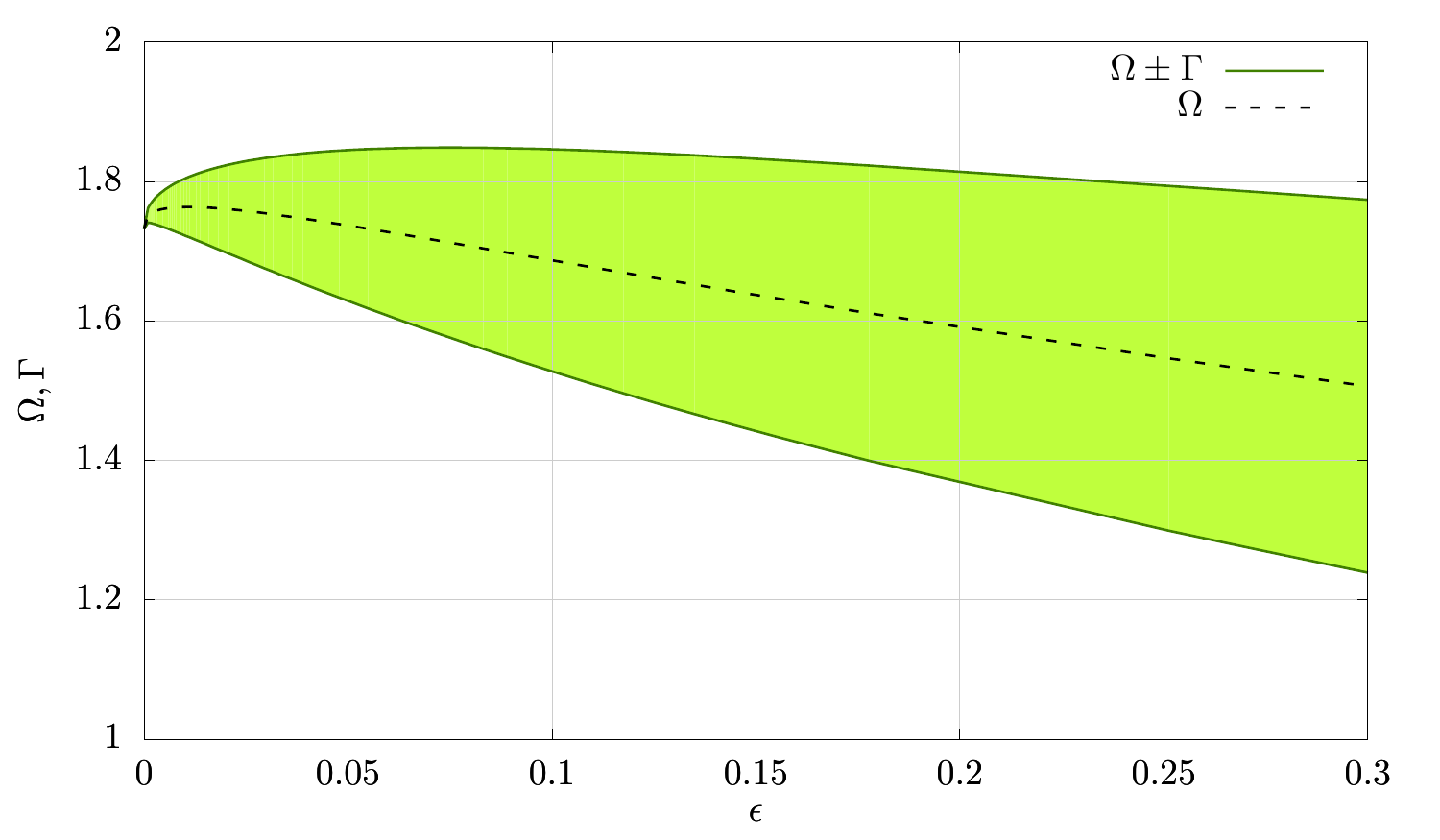}
 \caption{\small Quasinormal modes: frequency $\Omega$ and decay rate
$\Gamma$ vs $\epsilon$.}\label{fig:QNM}
\end{center}
\end{figure}

We found the frequencies of the quasinormal modes using one of
the simplest approaches, based on the Prony's method \cite{Prony}. 
We simulated the linear evolution with the excited profile of the $\phi^4$
oscillational mode and fitted a damped trigonometric function to the
field measured at the $x=0.88$ for times $5<t<30$. The results are
gathered in Table \ref{table:1:QNM} and shown in 
Fig.~\ref{fig:QNM}. It is possible that higher QNM exist, but only the
narrowest QNM contributes to the dynamics.

\begin{table}\centering
\caption{Quasinormal modes frequencies}\label{table:1:QNM}
\texttt{\small
\begin{tabular}{ccc}
  \hline
  \hline
 \hspace*{10mm}$\epsilon$\hspace*{10mm}& \hspace*{10mm}$\Omega$\hspace*{10mm} & \hspace*{10mm}$\Gamma$\hspace*{10mm} \\
  \hline
  0.000 & 1.73205 &0.00000 \\
  0.001 & 1.75152 &0.01049\\
0.002 & 1.75553 &0.01559\\
0.005 & 1.76099 &0.02678\\
0.010 & 1.76345 &0.04075\\
0.020 & 1.76079 &0.06229\\
0.030 & 1.75429 &0.07975\\
0.050 & 1.73697 &0.10813\\
0.100 & 1.68703 &0.15922\\
  \hline
\end{tabular}}
\end{table}
We found that a good fit to the resonance, valid for
$\epsilon<0.04$ with an accuracy of about 2\%, is
\begin{equation}
 \omega \approx 1.738 + 0.490\sqrt{\epsilon} -2.280\epsilon +
(0.325\sqrt{\epsilon}+0.783\epsilon)i\,.
\end{equation} 

\section{Numerical results}
\begin{figure}[!h]
 \begin{center}
 \includegraphics[width=1\linewidth]{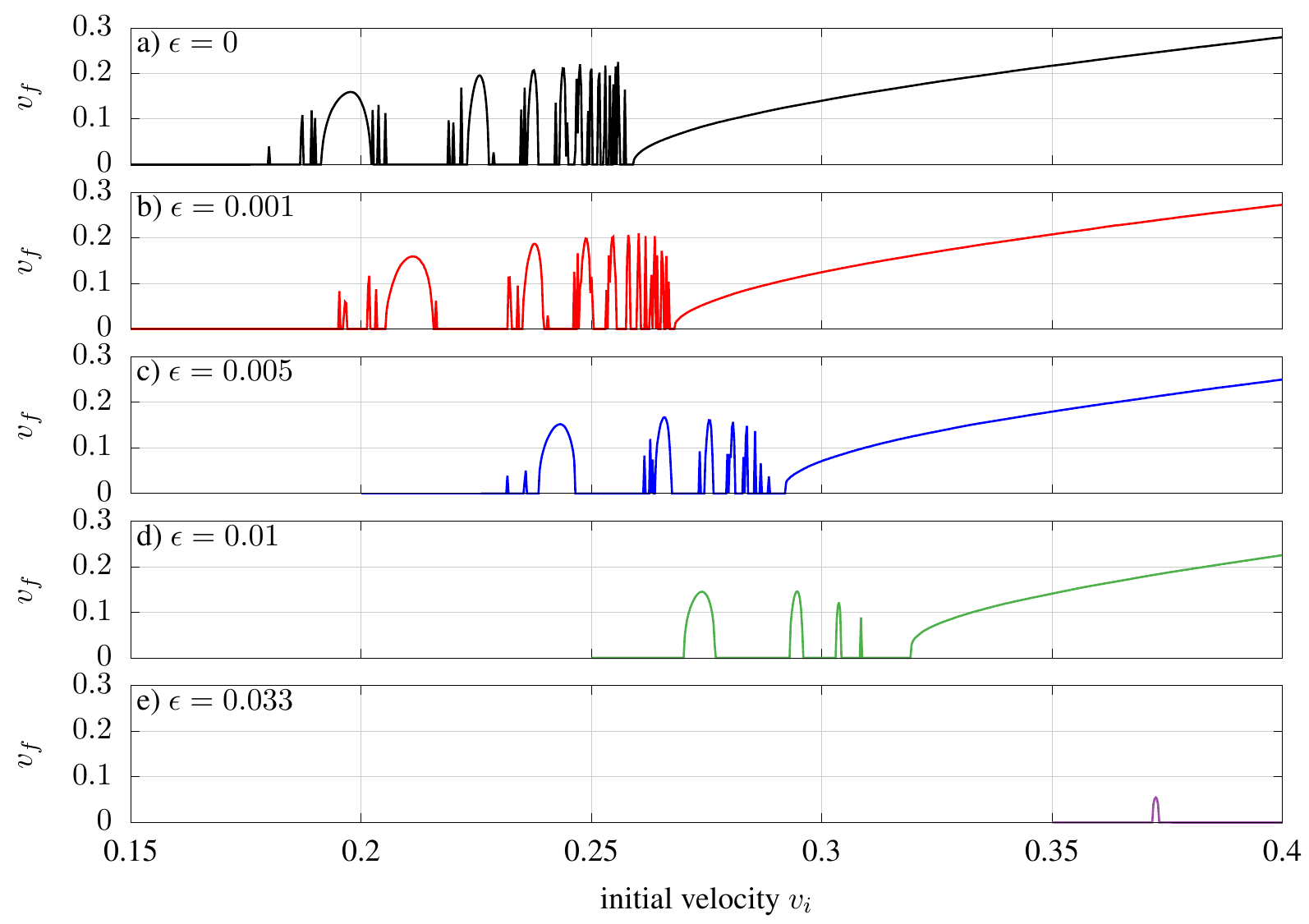}
 \caption{\small Final velocities of the escaping kink as a function
of initial velocity for different values of
$\epsilon$.}\label{fig::velocities}
\end{center}
\end{figure}

\begin{figure}
 \begin{center}
 \includegraphics[width=1\linewidth]{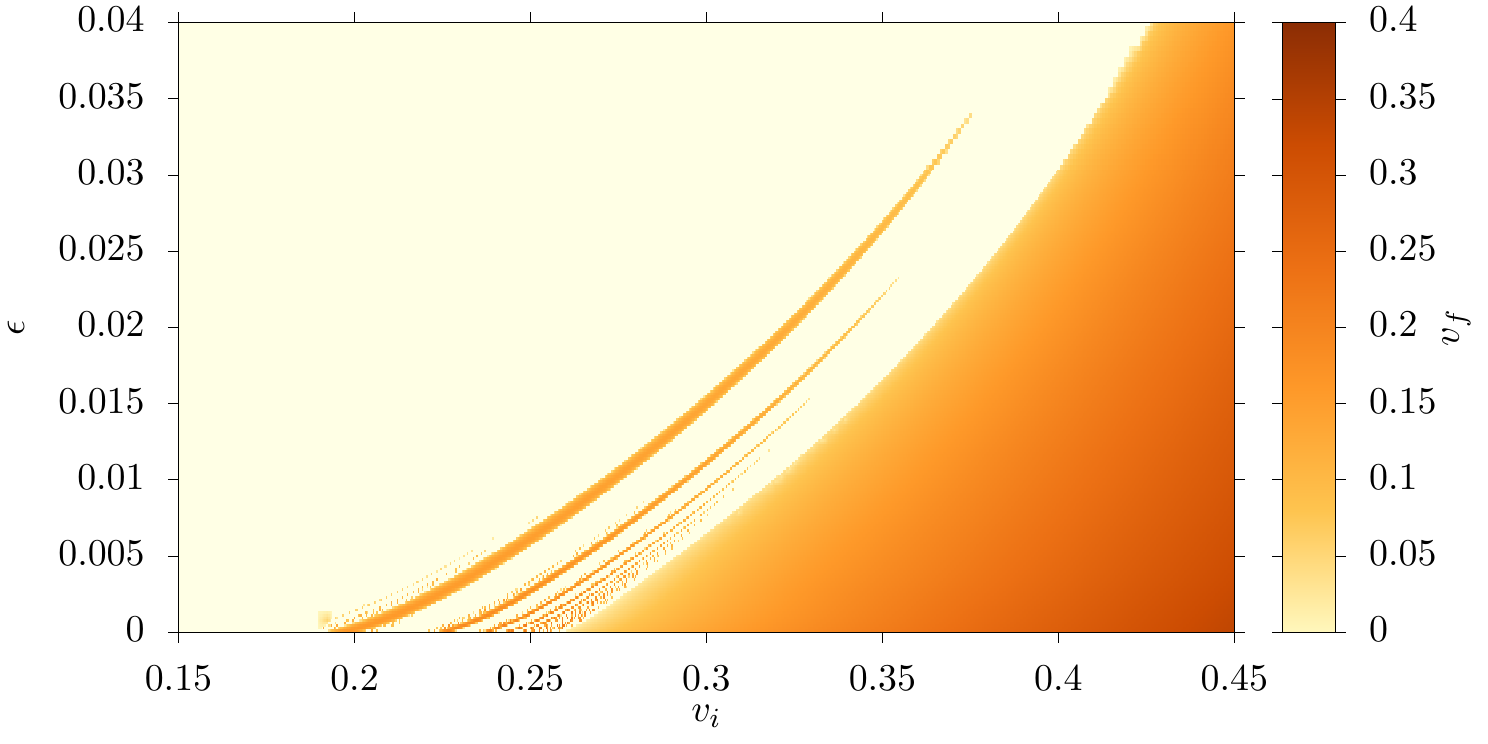}
 \caption{\small Scan of final velocities as a function of the
initial velocity $v_i$ and the perturbation parameter
$\epsilon$.}\label{fig::scan}
\end{center}
\end{figure}

For small values of $\epsilon$ the QNMs act very similarly to the
normal, oscillational mode (NM) of the $\phi^4$ model:  they store 
energy and via the resonant coupling they return it to the
translational modes of the kinks.  However not all the energy is given
back, as  some fraction of it escapes as the QNM decays exponentially.
The shorter the lifetime of the QNM, the more energy escapes, and
as a result bounce windows close. Our numerical simulations 
support this thesis, see Figures \ref{fig::velocities} and \ref{fig::scan}.
The bounce windows disappear, starting from those which for
$\epsilon=0$ had the largest
numbers of oscillations of the internal mode
and were more narrow. 
The windows vanish completely for  $\epsilon\approx 0.033$ which corresponds
to $\Gamma\approx 0.084$.

There is another interesting effect, worthy of further study.  
QNMs are very effective in
radiating the energy from the kinks.  For larger vales of $\epsilon$
one can observe that the critical velocity increases.  QNM are excited
but the lose energy very quickly, gluing the kink and antikink together.

We also expected that when the kinks annihilate they would form an oscillon. 
Surprisingly, this is not entirely true. 
A long-lived, almost periodic state is indeed created, but its fundamental frequency is above the  $m=1$ mass threshold of the deformed theory, though still below the 
$\phi^4$ mass threshold. 
As a result this object radiates via its first harmonic and decays faster than a standard oscillon. 
Initially its decay is slow and resembles that of the $\phi^4$ oscillon but in time, when the amplitude decreases, the decay rate
increases. We leave the detailed study of this object for future work.

\section{Effective model}
When the resonant fractal structure was found in the $\phi^4$ theory,
an effective model was proposed. 
The conjecture was that the field describing a
kink-antikink  collision should be described as a simple superposition of their profiles:
\begin{multline}\label{eq:effective1}
 \phi(x,t)\approx\phi_K(x-X)-\phi_K(x+X)+1+\\+
\sqrt{\frac32}A\left[\eta_d(x-X)-\eta_d(x+X)\right]
\end{multline} 
where $X(t)$ is the collective coordinate describing the position of
the kinks and $A(t)$ is the amplitude of the oscillational modes with
the profile $\eta_d(x)$. Substituting the above approximation into the
Lagrangian and integrating one obtains an effective (mechanical)
Lagrangian for $X$ and $A$.
Some further approximations were used, neglecting for example anharmonic terms.
After substituting (\ref{eq:potential1}) and integrating over spacial
dimension the Lagrangian could be written in the form
\begin{equation}
 L=a_1\dot X^2-a_2+a_3\dot A^2-a_4A^2+a_5A,
\end{equation} 
where the coefficients were given by appropriate integrals. 
This effective model gave a reasonably good approximation,
reconstructing qualitatively multibounce windows, and predicting a critical
velocity about 10\% 
higher than that observed in the solution of the full PDE. Originally
the coefficients (after appropriate rescaling) were given as
\begin{subequations}\label{eq:coeffs}
\begin{equation}\label{eq:a1}
 a_1(X)=\frac{4}{3}\left[1+\frac{6X\coth 2X-3}{\sinh^22X}\right],
\end{equation}
\begin{multline}
 a_2(X)=8\left[-\frac{2}{3}+
2X+3\coth 2X-\right.\\\left.-(2+6X)\coth^22X+4X\coth^32X\right],
\end{multline}
\begin{equation}
 a_3=1,\qquad a_4=3,
\end{equation} 
\begin{equation}\label{eq:wronga5}
 a_5(X)=-\sqrt{6}\pi\frac{\tanh^22X}{\cosh^22X}.
\end{equation}
\end{subequations}
Unfortunately, as pointed out in \cite{Weigel}, the effective model
had a typographic error which has been repeated in many of the following
papers. The 
term $a_5$ (in literature referred to as $F(X)$) should have a different form:
\begin{equation}
 a_5(X)=-3\pi\sqrt{\frac{3}{2}}
\left[2-2\tanh^3X-\frac{3}{\cosh^2X}+\frac{1}{\cosh^4X}\right].
\end{equation}
More surprisingly, the corrected version gave incorrect results,
including a prediction 
that the resonant structure would extend to all velocities.
The cure found in \cite{Weigel} was to take into account all the terms
without any
approximations. The full effective model had a further problem in that
the equations 
were singular for $X=0$, requiring an additional term to be introduced to
regularize the system of ODEs.

We decided to take the advantage of the simplicity of the
original model (\ref{eq:coeffs}) and treat it as some sort of toy or
phenomenological model. 
The appropriate equations of motion, neglecting higher terms in $A$,
can be written as
\begin{subequations}
\begin{equation}
\ddot{X} =\frac{1}{2a_1}\left(a_1'\dot X^2-a_2'+a_5'A\right),
\end{equation} 
\begin{equation}
\ddot{A} +\Omega^2A - \frac{1}{2}a_5=0
\end{equation} 
\end{subequations}
where the source term for $a_5$ is given by the equation (\ref{eq:wronga5}). 

\begin{figure}
 \begin{center}
 \includegraphics[width=1\linewidth]{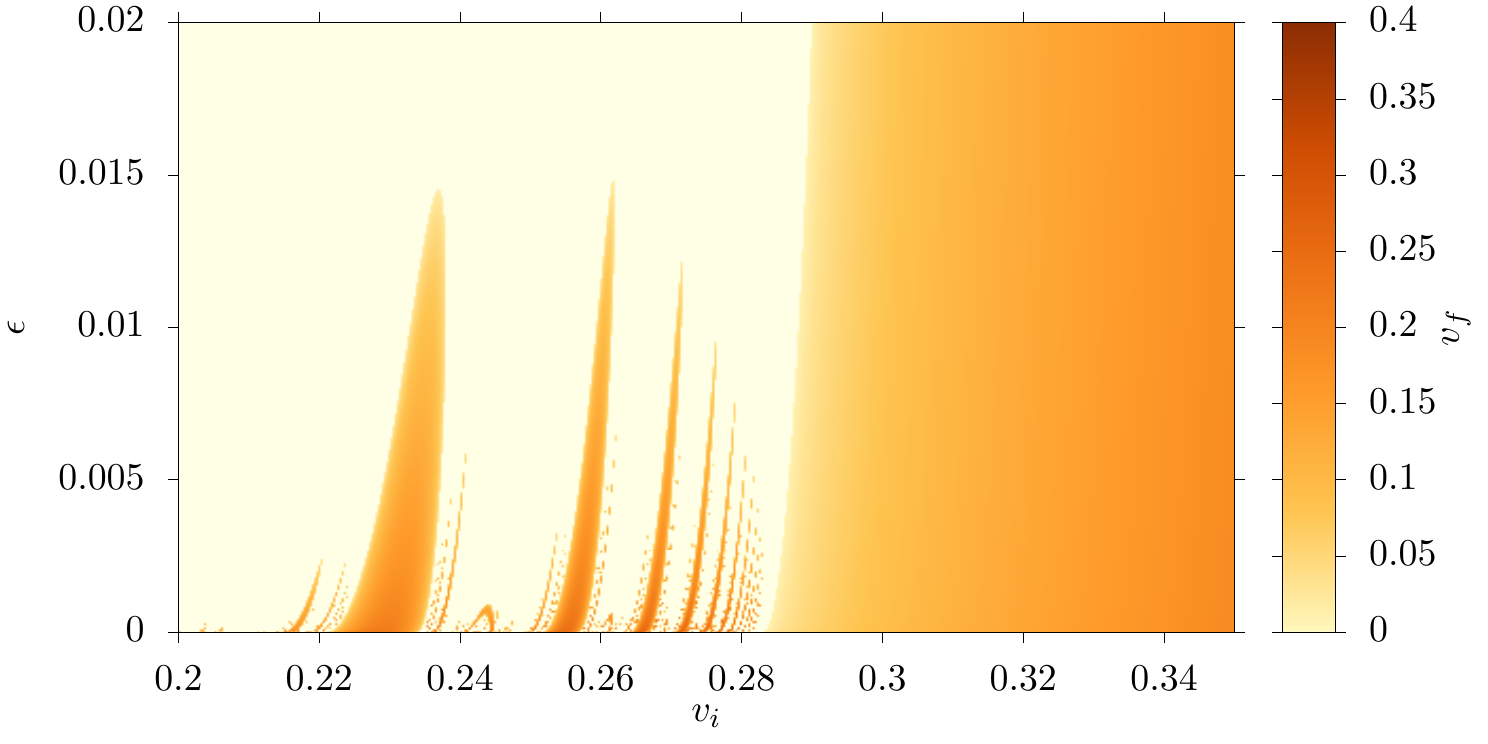}
 \caption{\small Scan of final velocities $v_f$ in the modified
effective model as a function of the initial velocity $v_i$ and the
perturbation 
parameter $\epsilon$. }\label{fig:scan2}
\end{center}
\end{figure}

We have adapted this method for our purposes. Our model differs very
little from $\phi^4$ for small values of $\epsilon$.  The asymptotic
profiles of the kinks are different but since in the original and our
model they decay exponentially fast, the actual profiles of the tails do
not contribute much to the collision process, especially for the large
velocities that we deal with.  The important modification is the
change in the nature of the oscillational mode which now becomes
the quasinormal mode.  We leave the frequency $\Omega\approx\omega_d$
but we add an additional damping term to the equation for $A(t)$
describing the decay of the mode
\begin{equation}
\ddot{A} + 2\Gamma\dot A +\Omega^2A - \frac{1}{2}a_5=0.
\end{equation}
The width of the QNM was approximated from the fit
$\Gamma\approx0.325\sqrt{\epsilon}$.
With this technique we have found that the value of $\epsilon$ for which
the resonant structure vanishes is $\epsilon_{cr}=0.015$, which is a
little less than half the value $\epsilon_{cr}=0.033$
found from the numerical solution of the 
full PDE. 
The results are shown in Figure \ref{fig:scan2}.
The resonance windows are again shifted towards higher
velocities, but these shifts are rather smaller in the effective model than in the full theory, as can be seen by comparing Figures \ref{fig:scan2} and \ref{fig::scan}.

Note that our modification does not include the change of the real
part of the frequency  of the QNM mode. 
Moreover the profile of the QNM is a complex function. Because of the
radiation tails, the QNM can interact at large distances. 
However, our modification qualitatively reproduces the results seen in
solutions of the full PDE, and we believe it will be a good
starting-point for further investigations.

\section{Comparison with other models}
To the best of our knowledge this is the first report concerning the
role of QNMs in kink collisions. One of our results is that the
resonant structure of bounces can be preserved for narrow resonances.
In addition we found that the critical frequency
increases as the width of the resonance becomes larger. This can be
explained by the fact
that the collision excites the QNM, which itself quickly gets rid off
energy. 
% The broader the resonance the shorter its half life time.
However it is also important that the collision excites the QNM.
Over the years, many models have been
studied in the context of colliding solitons, 
and it might be interesting to correlate the critical
velocities in other models with the existence of QNM.

It is known that neither sine-Gordon (sG) nor the $\phi^4$ models have QNM.
The linearised potentials for these two models are reflectionless
P\"oschl-Teller potentials. Moreover the sG model is integrable, and
there is no energy exchange between the scattering modes and the
solitons.  The critical velocity is exactly 0 -- the kinks always
separate after the collision.  For the $\phi^4$ model the critical
velocity is $v_{cr}=0.26$ but the first window opens for $v=0.18$. So
in a sense the oscillational mode prevents the kinks from separating for
$0.18<v<0.26$.

 Collisions in the  $\phi^6$ model also have a resonant structure.
However a different mechanism takes place in this model. For certain
kink configurations, when the vacuum with smaller mass lies between
the two kinks,
some separation dependent modes could be trapped between
the solitons.  On the contrary when the kinks collided with the
opposite arrangement
the collision did not show any
windows. However the critical frequency was more than six times
higher, $v_{cr}=0.289$. There are known solutions to the linearised
problem \cite{Lohe:1979mh}. To calculate the position of QNM it is
enough to know the asymptotic form of the solutions:
\begin{equation}
   \begin{cases}
      \eta_{+\infty}(x)\to e^{-ikx},\\
      \eta_{-\infty}(x)\to A(-q,-k)e^{-iqx}+A(q, -k)e^{iqx}
   \end{cases}
\end{equation} 
with
\begin{equation}
   \begin{gathered}
   q=\sqrt{\omega^2-1},\;k=\sqrt{\omega^2-4},\\
   A(q,k) =
\frac{\Gamma(1-ik)\Gamma(-iq)}{\Gamma(-\frac{1}{2}ik-
\frac{1}{2}iq+\frac{5}{2})\Gamma(-\frac{1}{2}ik-\frac{1}{2}iq-\frac{3}{2})}.
   \end{gathered}
\end{equation} 
The condition for the purely outgoing wave is that the reflection
coefficient $A(q,-k)/A(-q,-k)$ has a pole.  From the above form it is
straightforward to obtain the positions of the poles:
\begin{equation}
 \omega_n=2i\sqrt{\frac{(n+4)!}{(5+2n)^2n!}},\qquad n=0,1,2,\ldots
\end{equation}
Note that all of the poles are on the imaginary axis. These are not
unstable modes because the stability of the kink solution is
guaranteed by energy arguments and the
topological charge.

The nonintegrable double sine-Gordon model was studied in
\cite{Peyrard:1984qn}. For certain values of parameter $-1/4<\eta<0$
it was shown that no oscillational modes exist, and a resonant
structure was not found. Two critical velocities were reported:
$v_{cr}(\eta=-0.05)=0.112$ and $v_{cr}(\eta=-0.15)=0.390$. For the
value $\eta=-0.05$ there are no signs of QNM in the spectrum presented
in \cite{skyrmions}. If a QNM exists it is either very wide or is located
below the mass threshold. On the other hand,
for $\eta=-0.15$ there is a QNM with the
frequency $\omega=1.078+0.499i$, and it would be particularly
interesting to explore this case further in the light of the findings
of the present paper.

An important question regarding the problem of increasing the critical velocity is how much the quasinormal modes get excited.
If during collisions such  modes are only marginally excited, their contribution to the critical velocity would be negligible even if their life-times were large.

\section{Conclusions}
We have shown that the well-known fractal structure of multiple bounce
windows can be seen in models which neither have kinks with
oscillational modes,
nor kinks and antikinks forming  trapping potentials. 
Narrow quasinormal modes can play a similar role to the oscillational
modes.  There is however an important difference. QNM decay
exponentially, leading to energy leakage from the colliding kinks.
As the width of the QNM grows, the windows close, and we have found the
largest value of the QNM width which allowed for the formation of
a bounce window.
A further effect of the QNM is that the kinks need more kinetic
energy to separate due to the increased energy leakage, and
we observed that the critical velocity increased as the imaginary
part of the QNM frequency increased.
We have considered an effective model which differed from the standard
effective model for the $\phi^4$ theory by a damping term in the equation
describing the evolution of the mode, finding
that the value of the damping term corresponded well
with the width of the QNM for the upper limit allowing the bounce windows.
The additional term was responsible for the increase of the critical velocity.

It is worth mentioning that 
a very similar problem can be addressed in a two component model when
the $\phi^4$ field is coupled with a second 
second field with smaller mass. The oscillational mode from the
$\phi^4$ field can radiate through the second channel becoming a quasinormal
mode. 
A similar feature was pointed out in \cite{Forgacs:Monopole2004}.

In physics there are many objects which do not have bound or oscillational
modes but do have resonance modes. To name one example, the skyrmions
describing nucleons have resonances discovered by Roper \cite{Roper1964}. Such
resonances can also influence the process of soliton collision, and
the model that we have proposed in this paper could serve as a useful
toy example for such more-complicated situations.

\section*{Acknowledgments}
We would like to thank Piotr Bizo\'n for helpful discussions and
explanations.
Our research was supported in part by an
STFC consolidated grant ST/P000371/1 and in part by
the National Science Foundation
under Grant No.\ NSF PHY11-25915, and PED would like to thank 
KITP, Santa Barbara for the warm
hospitality as this paper was being finished.

\section*{References}

\bibliographystyle{elsarticle-num}
\bibliography{refs}

\end{document}